\documentclass[12pt, preprint]{aastex}

\begin{document}

 \title{Further Evidence that the Redshifts of AGN Galaxies May Contain Intrinsic Components}

\author{M.B. Bell\altaffilmark{1}}

\altaffiltext{1}{Herzberg Institute of Astrophysics,
National Research Council of Canada, 100 Sussex Drive, Ottawa,
ON, Canada K1A 0R6;
morley.bell@nrc.gc.ca}

\begin{abstract}

In the decreasing intrinsic redshift (DIR) model galaxies are assumed to be born as compact objects that have been ejected with large intrinsic redshift components, z$_{i}$, out of the nuclei of mature AGN galaxies. As young AGN galaxies (quasars) they are initially several magnitudes sub-luminous to mature galaxies but their luminosity gradually increases over 10$^{8}$ yrs, as z$_{i}$ decreases and they evolve into mature AGN galaxies (BLLacs, Seyferts and radio galaxies). Evidence presented here that on a log$z$-m$_{v}$ plot the bright edge of the AGN galaxy distribution at $z$ = 0.1 is unquestionably several magnitudes sub-luminous to the brightest radio galaxies is then strong support for this model and makes it likely that the high-redshift AGN galaxies (quasars) are also sub-luminous, having simply been pushed above the radio galaxies on a log$z$-m$_{v}$ plot by the presence of a large intrinsic component in their redshifts. An increase in luminosity below $z$ = 0.06 is also seen. It is associated in the DIR model with an increase in luminosity as the sources mature but, if real, is difficult to interpret in the cosmological redshift (CR) model since at this low redshift it is unlikely to be associated with a higher star formation rate or an increase in the material used to build galaxies. Whether it might be possible in the CR model to explain these results by selection effects is also examined.

\end{abstract}


\section{Introduction}
 
Because the belief that the redshift of quasars is cosmological has become so entrenched, and the consequences now of it being wrong are so enormous, astronomers are very reluctant to consider other possibilities. However,
there is increasing evidence that some galaxies may form around compact, seed objects ejected with a large intrinsic redshift component from the nuclei of mature active galaxies. In this model, as the intrinsic component decreases the compact objects evolve into mature active galaxies in a time frame of a few times $10^{8}$ yrs  \citep{arp97,arp98,arp99,bel02a,bel02b,bel02c,bel02d,bel04,bel06,bel06a,bel07,bur99,gal05,lop06}. In the DIR model radio galaxies represent the end of the AGN galaxy evolutionary sequence, where most of the intrinsic redshift component has disappeared and their luminosity has peaked. Only then can these objects be detected to large cosmological distances and can it be seen that they are good standard candles. There is every reason to assume that at each stage of their evolution (at each z$_{i}$ value) they will also be good standard candles. In this paper AGN refers to the active nucleus and AGN galaxy refers to the nucleus plus host galaxy.

It was recently demonstrated that the high redshift AGN galaxies detected to date appear to have a mean distance near 300 Mpc \citep{bel04}, and therefore few beyond $\sim500$ Mpc will have been detected.
However, in the DIR model it is assumed that this birthing process through compact object ejection has taken place at all cosmological epochs and that those galaxies that were born in the early universe still survive today, even though they will have almost certainly evolved beyond the mature AGN galaxy (radio galaxy) stage. Although they may no longer contain active nuclei, by this point in their evolution their redshifts will contain only a very small intrinsic redshift component. This remnant intrinsic redshift is observed to-day in common spiral galaxies \citep{tif96,tif97,bel03,bel04a}, and the local Hubble constant is found to be H$_{o}$ = 58 km s$^{-1}$ Mpc$^{-1}$ when the intrinsic components are removed \citep{bel03,bel04a}. This value is smaller than the value (H$_{o}$ = 72) obtained by \citet{fre01} before removal of the intrinsic components. In most respects the DIR model is perfectly compatible with the standard Big Bang model of the Universe. It differs mainly in the way galaxies are born and the claim that in this model at least the radio galaxies pass through an initial short-lived AGN period (10$^{8}$ yrs) in which their redshifts contain an intrinsic component that quickly disappears. After that, as they evolve through the next 10$^{10}$ years they can be used as they are today, to study cosmology. Although there is now a considerable amount of evidence supporting the DIR model, there are also some well-known arguments against this model that have been raised by those who support the CR model (e.g. the Lyman forest, lensing by intervening galaxies, etc.). An explanation of these arguments in the DIR model can be found in the Discussion section of a previous paper \citep{bel04}.

In the CR model the location of high-redshift AGN galaxies (quasars) on a log$z$-m$_{v}$ plot can be explained by the presence of a non-thermal component superimposed on their optical luminosity. In the DIR model their location on this plot is explained by the presence of a non-cosmological redshift component superimposed on their redshift. This paper uses an updated log$z$-m$_{v}$ plot containing over 100,000 AGN galaxies to compare the most luminous radio galaxies and first-ranked cluster galaxies at each redshift to the high luminosity edge of the AGN galaxy distribution in an attempt to see which model (CR or DIR model) can best explain the data. In this paper the standard candle (constant luminosity) slope is used as a reference to make luminosity comparisons at a given redshift. This is shown as a dashed line in Fig 1 and a solid line in Fig 2. Luminosity increases to the left.

\section{The Data}

A logz-m$_{v}$ plot for those radio sources with measured redshifts that were detected in the 1 Jy radio survey \citep{sti94} is presented in Fig 1. The quasars are plotted as filled circles and the radio galaxies as open squares. As discussed above, in the DIR model the radio galaxies are the objects that high-redshift quasars and other AGN galaxies evolve into when their intrinsic redshift component has largely disappeared. In Fig 1, first-ranked cluster galaxies \citep{san72a,kri78} are indicated by the dashed line. The most luminous radio galaxies, like first-ranked cluster galaxies, are clearly good standard candles to large cosmological distances, and their redshifts must then be cosmological, as expected in both the CR and DIR models since any intrinsic redshift component will have almost completely disappeared.
%The dashed line represents first-ranked cluster galaxies \citep{kri78,san72a}. 
%The dashed lines indicate the approximate locations of the intermediate and %high redshift AGN in the DIR model. There is every reason to assume that, like %the z$_{i} \sim 0$ radio galaxies, they are also good standard candles. If the %redshifts of the quasars are cosmological they cannot be standard candles. 

All the sources listed as quasars and active galaxies in the updated V$\acute{e}$ron-Cetty/V$\acute{e}$ron catalogue \citep{ver06} (hereafter VCVcat) are plotted in Fig 2. Since the VCVcat is made up of AGN galaxies from many different surveys, there will undoubtedly be differences in the selection criteria involved. However, since AGN galaxies are easily distinguishable from other types of galaxies, the normally strict selection criteria are not required in this case to obtain a source sample that is made up almost entirely of AGN galaxies. In that sense the VCVcat is probably the most complete sample of AGN galaxies available to-day. Because the source distribution in the plot in Fig 2 is \em continuous, \em the sources listed as quasars and AGN are clearly the same, and there is therefore no reason to separate them into two different categories as was done arbitrarily in the VCVcat. This should not be too surprising since they have long been lumped together in unification models \citep{ant93}. In Fig 2 the abrupt decrease in the number of sources for $0.5 < z < 3$ and m$_{v} > 21$ is explained by a faint magnitude cut-off near m$_{v}$ = 21m. It cannot affect the conclusions drawn here because at each redshift we are only comparing the bright, or high luminosity, edge of the source distribution (where the source density increases sharply when moving from bright to faint). For example, in Fig 2, at z = 0.03, 0.06, 0.15 and 1, the high luminosity edge of the AGN galaxy distribution is at m$_{v}$ = 14, 15, 18, 17, respectively.

However, some surveys have had other observer, or program-imposed limits applied that can also affect the bright edge of the source distribution and this is discussed in more detail in Section 3.1. The slope change in the high-redshift tail ($z > 3$) may be due to uncertainties in converting to visual magnitudes and/or to large k-dimming effects that have been unaccounted for. Whatever the cause, it will also not affect the arguments presented here that only apply to sources at lower redshifts.

\section{Discussion}

In Fig 1, the large triangle shows where the quasars would be located in the DIR model if the intrinsic component in their redshifts could be removed. All must lie below the radio galaxies. In this plot there are no AGN galaxies below the radio galaxies, and it is therefore easy to conclude that quasars are at the distance implied by their redshifts and are therefore super-luminous to first ranked cluster galaxies at all epochs. This was the conclusion drawn by \citet[see his Fig 4]{san72b} from a plot similar to Fig 1. Sandage argued that since no quasars lie to the right (fainter) of the radio galaxy distribution, this can be understood if a quasar consists of a normal, strong radio galaxy with a non-thermal component superimposed on its optical luminosity. He concluded from this evidence that quasars redshifts are cosmological.

In Fig 2 many of the high redshift quasars are also located above the radio galaxies, however, here most of the low- and intermediate-redshift AGN galaxies \em fall below the radio galaxy line. \em This is what is expected in the DIR model where AGN galaxies are born sub-luminous and reach their most luminous point when the intrinsic redshift component has disappeared. They must therefore all fall below the mature galaxy line. If those detected to date are all nearer than $\sim500$ Mpc \citep{bel04} most will also be located below the dashed line at z = 0.1 in Fig 2. This is what is seen in Fig 2 when the intrinsic component is small. \em The fact that low-redshift AGN galaxies are located below this line when the intrinsic component is too small to push them above it, suggests strongly that it is only the intrinsic component present in the high redshift sources that has pushed these sources above the radio galaxies. \em This argument is also supported by the shape of the plot in Fig 2, which starts out flat near z = 0.06, steepening gradually to z = 0.2 and then more rapidly to high redshifts. This conclusion is further supported by the fact that the $z_{i} \sim0$ AGN galaxies (radio galaxies) are good standard candles, and there is therefore no reason to think that the other AGN galaxies will not be, for a given intrinsic redshift value.
%Note that in the DIR model, although their luminosity increases with time as %they mature, and $z_{i}$ decreases, their mean luminosity is assumed not to %change with cosmological epoch. Unfortunately this cannot be proven because %intermediate and high intrinsic redshift sources are too faint to be detected %at large distances. However, the shape of the AGN galaxy distribution in Fig 2 %argues strongly that they contain a large intrinsic redshift component. Since %the location of the low-redshift sources in Fig 2 cannot be questioned (because %they have no significant intrinsic component to reposition them), their %location on the plot is also a confirmation of the distances estimated %previously \citep{bel04} for these sources (less than $\sim500$ Mpc).

%In the CR model, the fact that AGN start out super-luminous at high redshifts, %become sub-luminous at intermediate and low redshifts, and then increase to %luminosities near those of first-ranked cluster galaxies at low redshifts is %not easily explained because no luminosity variations are seen in similar types %of objects that occupy the same space in the Universe. 

Because almost all of the AGN galaxies are less luminous than the highest luminosity radio galaxies and first-ranked cluster galaxies at redshifts below z $\sim0.3$, the explanation proposed by \citet{san72b} can no longer be valid. Quasars cannot be normal radio galaxies, or even Seyferts, with a non-thermal optical component superimposed. In fact, since the high luminosity edge of the AGN galaxy distribution in Fig 2 is $\sim3$ mag fainter than the high luminosity edge of the radio galaxies at z = 0.1, if quasars are sub-luminous galaxies brightened by a superimposed non-thermal optical component, at z = 2 this superimposed component would have to increase the optical luminosity of the source by up to $\sim9$ magnitudes. This could even get worse at higher redshifts when k-dimming effects are included, which would make the standard model involving a superimposed non-thermal nuclear component increasingly difficult to believe.

In the CR model the peak in quasar activity (luminosity and number) near z = 2 is assumed to be associated with a period when the star formation rate was higher than at present, and because there was more raw material around to make galaxies. In Fig 2, not only does the high luminosity edge of the AGN galaxies get intrinsically much fainter towards low redshifts (moving further to the right relative to the standard candle slope), below $z \sim0.3$ this decrease in luminosity begins to slow down. \em Below z = 0.1 their luminosity begins to increase again, eventually approaching that of the brightest radio galaxies. \em How is this to be explained in the CR model when we can no longer use the argument that there is more raw material around? This is one of the questions that will need to be addressed if the CR model is to continue to be favored, since this increase is exactly what is predicted in the DIR model as the AGN galaxies mature into radio galaxies. One possible explanation in the CR model is discussed in the following section. 

\subsection{Selection Effects in the Data}

Although in a sample like VCVcat it is difficult to take into account all of the selection effects that might be active, since the Sloan Digital Sky Survey (SDSS) sources are likely to make up the largest single portion of the sample the target selection process in that survey is worth examining. First, the survey is sensitive to all redshifts lower than $z$ = 5.8, and the overall completeness is expected to be over $90\%$ \citep{ric02}.  Extended sources were also targeted as low-redshift quasar candidates in order to investigate the evolution of AGN at the faint end of the luminosity function. During the color selection process no distinction was made between quasars and the less luminous Seyfert nuclei. Objects that had the colors of low-redshift AGN galaxies were targeted even if they were resolved. This policy was in contrast to some other quasar surveys that reject extended objects, thereby imposing a lower limit to the redshift distribution of the survey \citep{ric02}. In addition to selecting \em normal \em quasars, the selection algorithm also makes it sensitive to atypical AGN such as broad absorption line quasars and heavily reddened quasars \citep{ric02}. 

%It thus seems apparent that the SDSS program made every effort possible to %ensure that few AGN galaxies were missed.

In addition to the detection limit set by the sensitivity of the observing system the SDSS also contains two additional observer, or program-imposed, limits. One of these was a faint-edge limit at $i^{*}$ = 19.1m, and the other was a bright-edge cut-off at $i^{*}$ = 15m. The reasons why these limits were imposed can be found in \citet{ric02}. 
Although color-selected quasar candidates below z = 3 were only targeted to a Galactic extinction-corrected $i^{*}$ magnitude of 19.1, as noted above, since we are only examining the bright edge of the log$z$-m$_{v}$ plot, this faint edge limit is not expected to have affected the results. However, the bright edge cut-off at $i^{*}$ = 15m could have affected the shape of the bright edge of the log$z$-m$_{v}$ plot and this needs to be examined more closely.

In Fig 2, for $0.7 < z <  3$ it is possible that the bright edge cut-off could have prevented the detection of some of the brighter sources, although if many were missed we might expect to see some evidence of a sharp cut-off along the bright edge similar to that seen at m$_{v} \sim21$m. None is seen. Furthermore, since the bright edge of the distribution between $z$ = 0.1 and $z$ = 0.5 is at least 1 magnitude fainter than many sources detected at the higher redshifts it seems unlikely that the $i^{*}$ = 15 limit could have significantly affected the bright edge of the distribution in this redshift range. In fact, it is apparent from Fig 2 of \citet{sch07} (which is a plot of the $i$ magnitude of the 77,429 objects in the SDSS Fifth Data Release quasar catalogue versus redshift) that in the SDSS catalogue it is unlikely that many sources were missed at any redshift because of the cut-off at $i$ = 15.

It is also worth noting that the sources that lie outside the limits imposed in the SDSS have not been discarded. SDSS photometry for those objects brighter than $i^{*}$ = 15 is sufficiently accurate that they can be used in follow-up studies should the need arise. \citep{ric02}.

In Fig 2 there is also an increase seen in the number of AGN galaxies as $z$ increases. Such an increase is expected in the CR model where the redshift is distance related and where it would be due to the increasing volume of space sampled as $z$ increases. This would then support the CR model. However, it needs to be kept in mind also that if a bright edge cut-off is affecting the shape of the bright edge of the source distribution, it would presumably also have created this increase in source number with redshift by preventing the detection of many more of the bright sources at low redshifts. In the DIR model, where the redshift of AGN galaxies is age related, the number density of sources as a function of cosmological redshift can only be determined after the intrinsic component is removed.

This paper examines the AGN galaxies listed in the VCVcat and draws conclusions based on that sample. It contains the quasars found in the SDSS that were available at the time the catalog was prepared, and approximately 11,000 Seyferts and BLLacs, but whether the current VCVcat contains many AGN galaxies found in the SDSS galaxy survey is unclear.

\citet{hao05a,hao05b} have pointed out that although the color selection technique used in the SDSS is very efficient, selecting AGN galaxies is a complex process and requires that the optical luminosity of the active nucleus be at least comparable to the luminosity of the host galaxy for the color to be distinctive. Thus the color selection systematically misses AGN galaxies with less luminous nuclei at low redshift. If mainly faint sources at redshifts below z = 0.08 were missed, it is conceivable that the bright edge currently visible near m$_{v} \sim14$ might have been created by the selection process. In this case there might be no luminosity increase below z = 0.1, which would be more easily explained in the CR model. However, if the VCVcat does not contain many AGN galaxies found in the SDSS galaxy survey this would not be a problem here. Furthermore, in the DIR model, where the luminosity of the host galaxy is predicted to increase as it matures, presumably bright AGN galaxies as well as faint ones could be missed if the host galaxy has brightened significantly so as to swamp the nucleus. Also, in Fig 2, the bend in the distribution towards higher luminosities near z = 0.07 and $m_{v}$ = 18 does seem to point to a real increase in the luminosity at lower redshifts. However, if this sample is incomplete at low redshifts for a particular magnitude range, the conclusions drawn here may change when a more complete sample becomes available. 
 
Also, if AGN galaxies at vastly different redshifts are to be compared, as here, it is important that the optical magnitude of the entire galaxy be used and not simply that of the nucleus. It is the total magnitude that has invariably been used for high redshift quasars because of the difficulty of separating the nuclear and host contributions. \citet{hao05b} point out that, in attempting to obtain the luminosity function of the active nucleus, it is important that it not be contaminated by the host galaxy. Since the brightening predicted in the DIR model below z = 0.1 is due to the host galaxy maturing and increasing in luminosity, the contribution from the entire host galaxy must be included in the magnitudes used in the logz-m$_{v}$ plot if the brightening is to be detected. Although the luminosity of the nucleus may be adequate in determining the luminosity function of the active nucleus in the CR model, because of the complex process required to identify AGN galaxies \citep{hao05b}, obviously great care will be required in obtaining the magnitudes of low redshift AGN galaxies if they are to be used in logz-m$_{v}$ plots.

\section{Conclusion}

The most luminous radio galaxies and first-ranked cluster galaxies have been compared here to the high luminosity edge of the AGN galaxy distribution on a log$z$-m$_{v}$ plot. It is found that while the radio galaxies and cluster galaxies are good standard candles at all epochs, the luminosity of the AGN galaxies varies significantly from one epoch to another. Compared to the comparison galaxies the AGN galaxies are found to be super-luminous at high redshifts, but become sub-luminous as the redshift decreases. These new results show that below $z $ = 0.3 the rate of luminosity decrease begins to slow down and below $z$ = 0.1 the luminosity begins to increase again. Although their apparent super-luminous nature at high z can be explained by a higher star formation rate, and the fact that there might have been more raw material around to make galaxies at that epoch, a luminosity increase below $z$ = 0.1 is more difficult to explain when these arguments are unlikely explanations. It is therefore concluded here that the evidence favors the argument that the high redshift AGN galaxies (quasars) that do lie above the mature galaxy line on a log$z$-m$_{v}$ plot have all been pushed there because of a large intrinsic component in their redshifts and not because they have a superimposed non-thermal component that is many magnitudes brighter than that seen in radio galaxies. All AGN galaxies then will be sub-luminous to mature galaxies, as predicted in the DIR model. For a given intrinsic redshift component, all are likely also to be good standard candles. Finally, if it turns out that many faint AGN galaxies at low redshifts have been missed in a particular magnitude range the conclusion that the bright edge of the logz-m$_{v}$ plot increases again in this redshift range may need to be re-evaluated. Such an effect might be introduced by the selection effect discussed by \citet{hao05a,hao05b}, but only if the VCVcat contains many of the SDSS AGN galaxies, as explained in Sec 3.1.

I wish to thank two anonymous referees for suggestions on how this paper might be improved. I also thank Dr. D. McDiarmid for helpful comments.

%\end{document}

\clearpage

\begin{figure}
\hspace{-1.0cm}
\vspace{-1.5cm}
\epsscale{0.9}
\plotone{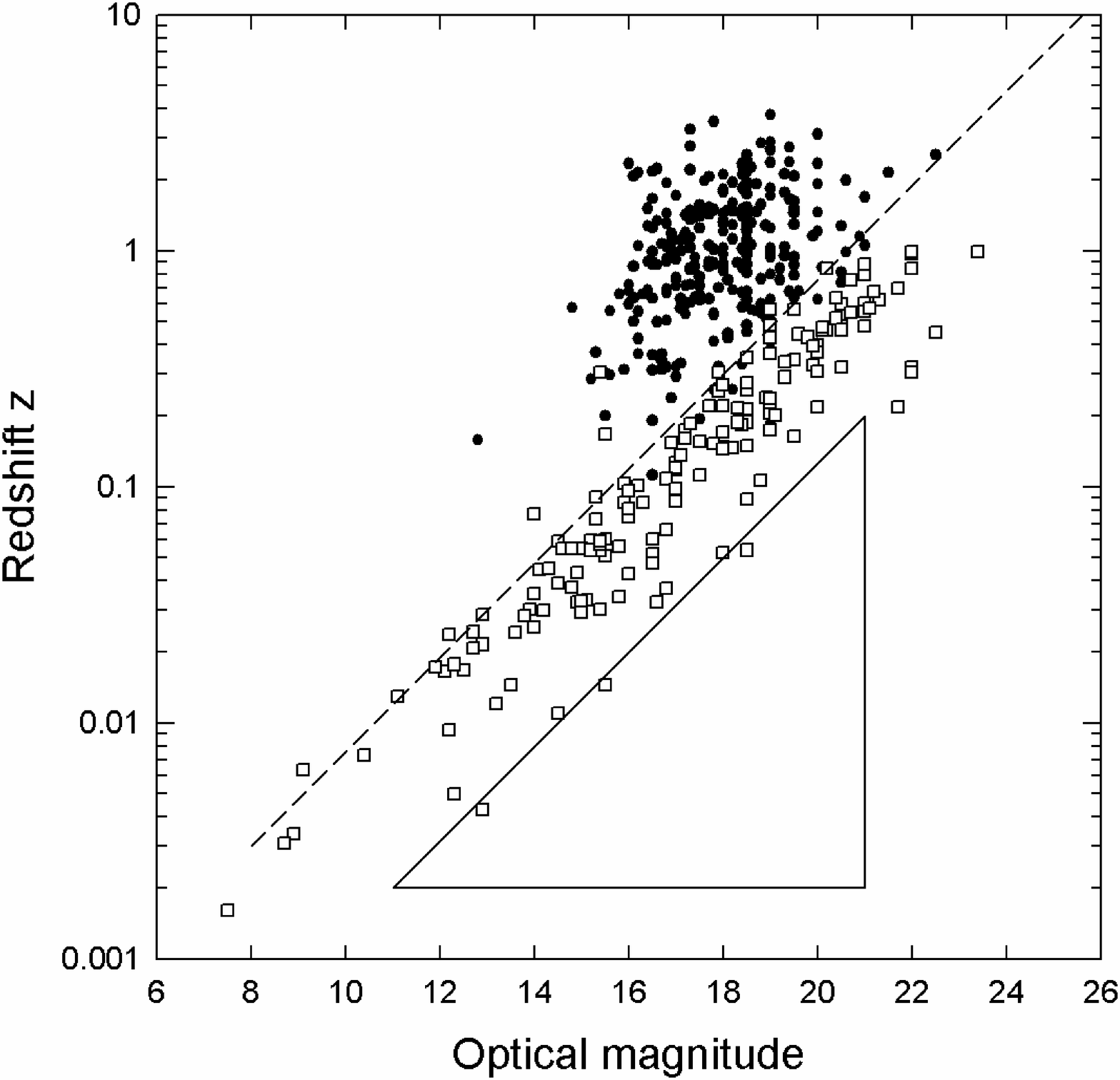}
%\plotone{stickel1.eps}
\caption{{Plot of redshift versus optical magnitude for quasars (filled circles) and radio galaxies (open squares) from \citep{sti94}. The dashed line represents brightest cluster galaxies from \citep{san72a,kri78}. See text for an explanation of the triangle. \label{fig1}}}
\end{figure}

\clearpage

\begin{figure}
\hspace{-1.0cm}
\vspace{-0.5cm}
\epsscale{0.9}
\plotone{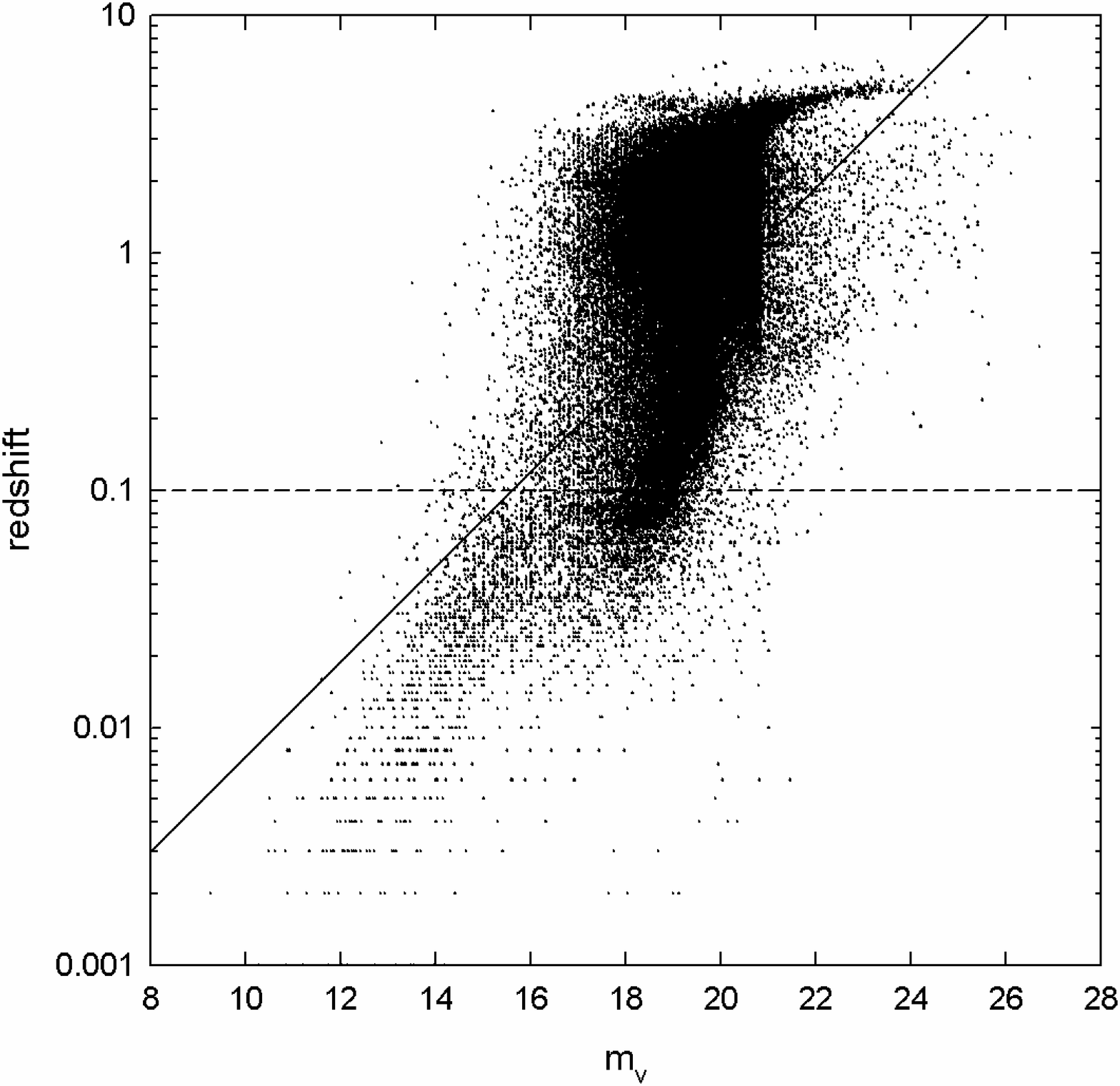}
%\plotone{qsoall20ptf2.eps}
%\plotone{qso10agnall12point.eps}
\caption{{Log$z$-m$_{v}$ plot of all 106,958 sources listed as quasars and active galaxies in the V$\acute{e}$ron-Cetty-V$\acute{e}$ron catalogue. The solid line indicates first-ranked clusters from \citet{san72a,kri78}. The dashed line indicates the maximum distance for high-redshift AGN detected to date from \citep{bel04}. In the DIR model any AGN that lie above this line have been pushed there by the presence of an intrinsic redshift component. \label{fig2}}}
\end{figure}
 
\end{document}